\documentclass{elsart}

\input{epsf}

\newcommand{\bfc}{{\bf c}}
\newcommand{\bfE}{{\bf E}}
\newcommand{\bfJ}{{\bf J}}
\newcommand{\bfS}{{\bf S}}
\newcommand{\bfu}{{\bf u}}
\newcommand{\bfV}{{\bf V}}
\newcommand{\bfx}{{\bf x}}
\newcommand{\bfy}{{\bf y}}
\newcommand{\calL}{{\bf L}}
\newcommand{\calN}{{\bf N}}
\newcommand{\bfnabla}{{\mbox{\boldmath $\nabla$}}}
\newcommand{\mychoose}[2]{{{#1}\choose{#2}}}

\begin{document}
\begin{frontmatter}
\title{Lattice Gases and Cellular Automata}
\author{Bruce M. Boghosian}
\address{
{Center for Computational Science, Boston University, 3 Cummington Street,}\\
{Boston, MA 02215, U.S.A.}\\[0.15truein]
{Electronic mail: {\tt bruceb@bu.edu}}\\
{Web page: {\tt http://physics.bu.edu/$\sim$bruceb}}
}
\begin{abstract}
We review the class of cellular automata known as {\sl lattice gases},
and their applications to problems in physics and materials science.
The presentation is self-contained, and assumes very little prior
knowledge of the subject.  Hydrodynamic lattice gases are emphasized,
and non-lattice-gas cellular automata -- even those with physical
applications -- are not treated at all.  We begin with a review of
lattice gases as the term is understood in equilibrium statistical
physics.  We then discuss the various methods that have been proposed to
simulate hydrodynamics with a lattice gas, leading up to the 1986
discovery of a lattice gas for the isotropic Navier-Stokes equations.
Finally, we discuss variants of lattice-gas models that have been used
for the simulation of complex fluids.
\end{abstract}
\begin{keyword}
Cellular automata, lattice gases, hydrodynamics, discrete kinetic
theory, Ising model, complex fluids, microemulsions
\end{keyword}
\end{frontmatter}

\section{Historical Background}
\subsection{The Ising Model}

\begin{figure}
  \center{\mbox{\epsfxsize=3.0truein\epsffile{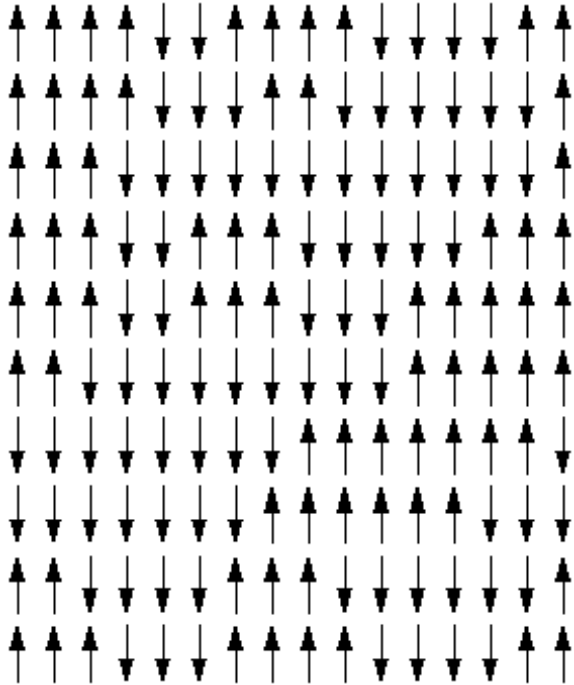}}}
  \caption{{\bf $D=2$ Ising model:}  At each lattice point there is a
  spin, represented here by an arrow, pointing either up or down.}
  \label{fig:ising}
\end{figure}

The use of lattice gases for the study of equilibrium statistical
mechanics dates back to a 1920 paper of Lenz~\cite{bib:lenz} in which he
proposed to model a ferromagnet by a regular $D$-dimensional lattice
$\calL$ of two-state ``spins.''  Physically, these may be thought of as
the magnetization vectors of elemental magnetic domains, and the model
constrains them to point in one of two directions, say ``up'' and
``down.''  For a two-dimensional lattice, this is illustrated
schematically in Fig.~\ref{fig:ising}.  Mathematically, the state of the
system can be described by the collection of variables $S(\bfx)$,
indexed by the lattice points $\bfx\in\calL$, and taking their values
from the set $\{-1,+1\}$; here $S(\bfx)=+1$ means that the spin at site
$\bfx$ is pointing up, and $S(\bfx)=-1$ means that it is pointing down.
If we suppose that the lattice has a total of $N\equiv |\calL|$ sites,
then the total number of possible states of the system is $2^N$

To use these spins as a model of ferromagnetism, it was necessary to
assign an energy to each of these $2^N$ states, in such a way as to make
it energetically favorable for each spin to align with an externally
applied magnetic field $\alpha$, and for neighboring spins to align with
each other.  The first of these goals is achieved by including an energy
contribution $\alpha S(\bfx)$ for each spin present, and the second by
including an energy contribution $JS(\bfx)S(\bfy)$ for each pair of
neighboring sites $\bfx$ and $\bfy$.  Thus, the full energy of the
system is
\[
H(\bfS) =
\alpha\sum_\bfx S(\bfx)-
\frac{J}{2}\sum_\bfx\sum_{\bfy\in\calN(\bfx)} S(\bfx)S(\bfy),
\]
where $\calN(\bfx)$ denotes the set of sites neighboring site $\bfx$,
and the factor of $1/2$ in front of the second term prevents
double-counting of the pairs of spins.

To use this to study the equilibrium properties of a ferromagnet, it is
necessary to compute the partition function
\[
Z(K, h)\equiv\lim_{N\rightarrow\infty}
\sum_\bfS \exp \left[-\frac{H(\bfS)}{k_BT}\right],
\]
where $T$ is the temperature, $K\equiv J/(k_BT)$,
$h\equiv\alpha/(k_BT)$, the sum over $\bfS$ includes all $2^N$ possible
states of the system, and we have taken the {\em thermodynamic limit\/}
by letting the number of spins go to infinity.

Lenz posed the problem of calculating this quantity to his student
Ising, who solved it for a one-dimensional lattice of spins in
1925~\cite{bib:ising}.  While Ising's $D=1$ solution is elementary,
Onsager's $D=2$ solution for $h=0$ required almost another twenty
years~\cite{bib:onsager} to complete, and is significantly more
complicated.  The solution for the critical exponents for $D=2$ with
$h\neq 0$ is a much more recent development, first published by
Zamalodchikov~\cite{bib:zama} in 1989.  The problem for $D=3$ is
outstanding, even for $h=0$.

\subsection{Universality and Materials Science}

One might wonder why so much effort has been devoted to the Ising model
when it is clearly only a crude idealization of a real ferromagnet.
Certainly, nobody expects the detailed functional form of, say, the
dependence of the Ising model's magnetization
\[
M(K, h) =
\frac{\sum_\bfx^N S(\bfx)\exp\left[-\frac{H(\bfx)}{k_BT}\right]}
     {\sum_\bfx^N        \exp\left[-\frac{H(\bfx)}{k_BT}\right]} =
\frac{\partial\ln Z(K, h)}{\partial h}
\]
on the temperature $T$ to be valid for any real material.  There are,
however, good reasons to believe that certain features of this
functional form are {\em universal\/} -- that is, model-independent.
This is particularly true near criticality (in the $D=2$ and $D=3$ Ising
models), where the spin-spin correlation length diverges, and
fluctuations at all length scales are present.  For example, at zero
applied field and near criticality, the magnetization varies as
\[
M =
\left\{
\begin{array}{ll}
0 & \mbox{for $T>T_c$}\\
M_0 (\frac{T_c - T}{T_c})^\beta & \mbox{for $T\leq T_c$},
\end{array}
\right.
\]
where $T_c$ is the critical temperature, $M_0$ is a proportionality
constant, and $\beta$ is an example of what is called a {\em critical
exponent}.  The scale invariance of the fluctuations at the critical
point allow a renormalization group treatment which indicates that the
critical exponent should be rather insensitive to the particular model
Hamiltonian used.

In fact, critical exponents should depend on only the dimensionality of
the space and the symmetries of the underlying Hamiltonian function.
For example, the unmagnetized Ising-model Hamiltonian is invariant under
the symmetry group $Z_2$ -- that is, multiplication in the set
$\{-1,+1\}$ -- because the energy is invariant under the inversion of
all the spins in the system.  Systems with $Z_2$ symmetry are expected
to have $\beta=1/8$ in $D=2$, and $\beta\approx 0.33$ in $D=3$.  A
related lattice spin model, called the Heisenberg model, endows each
spin with a vector orientation in three dimensions and has an
interaction Hamiltonian that depends only on dot products of these
vectors at neighboring sites.  Since these are invariant under the
continuous group of SO(3) rotations, we might expect a different
critical exponent for $\beta$, and in fact this is the case:
$\beta\approx 0.36$ for the $D=3$ Heisenberg model.

Thus, universality teaches us that it is possible to learn some ``real
physics'' by studying highly idealized models such as the Ising model.
This realization led to a flurry of variants of lattice spin models,
appropriate to various real materials.  As an example, we consider an
ingenious model developed by Widom~\cite{bib:widom} to describe
microemulsions.  A microemulsion is created by the addition of a {\em
surfactant\/} or {\em amphiphile\/} to a mixture of two immiscible
fluids, such as oil and water.  An amphiphile is a chemical that
typically has an ionic {\em hydrophyllic\/} end that likes to sit in
water, covalently bonded to a hydrocarbon chain which is {\em
hydrophobic\/} in that it prefers to sit in oil.  This situation gives
rise to two crucial properties: First, the free energy of the amphiphile
is lowest when it lives on the interface between the two fluids.
Second, the presence of the amphiphiles on the interface gives it a
rigidity, or {\em bending energy}, that is proportional to the square of
the local mean curvature~\footnote{More generally, it is proportional to
the square of the difference of the local mean curvature and some
spontaneous value thereof. Note that this rigidity is in addition to the
surface tension that is always present between immiscible fluids.}.
Together, these two properties can give rise to some spectacular
behavior.  For example, thanks to the first property, when there is
insufficient interface to accomodate the amphiphile, it becomes
energetically profitable for the amphiphile to create new interface to
inhabit.  It does this by breaking up the separated mixture of water and
oil into emulsion droplets.  The droplets become smaller as more
amphiphile is added, until they become so small (on the order of 50 nm)
that the curvature energy mentioned in the second property removes the
incentive for them to get any smaller.  If the amount of surfactant
continues to increase past this point a sponge phase can result, or, at
lower temperatures, ordered {\em lyotropic\/} phases consisting of
alternating sheets or tubes of oil and water.  The self-organization of
these structures as a result of relatively simple chemical properties is
a methodology for nanofabrication, and scientific and industrial uses of
these materials abound.

\begin{figure}
  \center{\mbox{\epsfxsize=3.0truein\epsffile{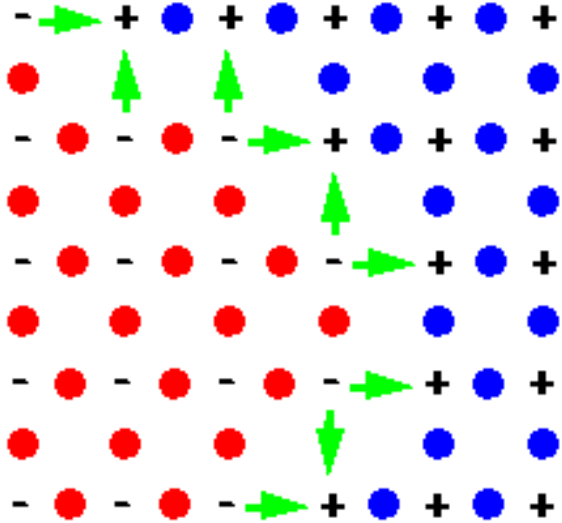}}}
  \caption{{\bf Widom's microemulsion model:} Edges connecting two
  ``up'' spins contain water, and those connecting two ``down'' spins
  contain oil.  Edges connecting an ``up'' spin with a ``down'' spin
  contain amphiphile, with hydrophyllic end oriented toward the ``up''
  spin.}
\label{fig:widom}
\end{figure}

To model microemulsions using a lattice gas, Widom~\cite{bib:widom}
introduced, in the mid 1980's, a model very similar to that of Ising, in
that each site on a Cartesian grid could be in one of two states,
$S(\bfx)\in\{-1,+1\}$.  One clever innovation of this model is that it
situates the particles on the links between the lattice vertices, rather
than on the vertices themselves.  Links can be characterized by the
values of $S$ at the vertices they connect.  Thus, each link can be in
any one of four possible states: Links that connect two $S=0$ vertices
are said to contain water.  Links that connect two $S=1$ vertices are
said to contain oil.  Finally, links that begin at an $S=1$ ($S=0$)
vertex and end at an $S=0$ ($S=1$) vertex are said to contain amphiphile
whose hydrophyllic end points toward the end (beginning) of the link.
The great advantage of this model is that the representation itself
literally forces the amphiphile to live on the interface between the oil
and water, and orients it correctly.  This is illustrated in
Fig.~\ref{fig:widom}.

The Hamiltonian for Widom's model includes two-spin interactions, as
were present in the Ising model Hamiltonian, but it also includes
three-spin interactions, necessary to capture the curvature energy
described above and required for the rigidity of the interface.  Since
the number of particles of each type may change when a spin is flipped,
it is also necessary to append terms of the form $\mu_W n_W+\mu_O
n_O+\mu_A n_A$, where the $\mu$'s are the (fixed) chemical potential per
particle and the $n$'s are the particle numbers of each species, and run
the Monte Carlo simulation in the grand canonical ensemble.

Widom's model has been used to reproduce much of the above-described
phenomenology associated with amphiphilic fluids.  In particular,
droplet, sponge and lyotropic phases are all seen, as are coexistence
regions between these phases.

\subsection{From Statics to Dynamics}

The Ising model is discussed in many (perhaps most) statistical physics
textbooks.  Our purpose for bringing it up in this context is merely to
note that it has two ``false identities.''  First, it looks a bit like a
{\em Hamiltonian dynamical system\/} because it has an energy function
(which we have obligingly denoted by $H$) that is exponentiated and
integrated to get a partition function -- just as is done with the
Hamiltonian of a system of classical particles when deriving, e.g., the
Mayer cluster expansion.  Second, it looks a bit like a {\em cellular
automaton\/} (CA) because its state can be represented by a single bit
at each site.  In fact, as defined above, the Ising model is neither of
these things.  Indeed, it is not a dynamical system at all.  It is posed
only as an {\em equilibrium\/} model.  You can use it to find a
partition function, and from that derive all of its thermodynamic
properties, but thermodynamics (its name notwithstanding) is not
dynamics at all.  When it is used to study the difference between two
states of a given system, it pays little or no attention to the
particulars of the dynamical process that takes the system between those
states.  By contrast, Hamiltonian dynamical systems require a continuous
phase space, endowed with a Poisson bracket that is antisymmetric and
obeys the Jacobi identity; given such a bracket structure, any phase
function $H$ defines a deterministic dynamics on the phase space.
Likewise, a CA is a Markovian dynamical system on a discrete phase
space, with transition rules that are applied to each site
simultaneously and depend only upon the states of neighboring sites.

When the Ising model is simulated on a computer to study its equilibrium
properties, it begins to look more like a dynamical system.  The usual
method is to perform a Monte Carlo simulation by flipping spins
randomly, and then accepting or rejecting these flips according to the
Metropolis algorithm~\cite{bib:mmc}, while sampling desired equilibrium
properties.  Because the Metropolis algorithm is Markovian in nature,
this is a dynamical system in the mathematical sense of the word.
However it is still neither Hamiltonian nor a CA.  The discrete phase
space would seem to distinguish it from most Hamiltonian
systems~\footnote{There have been interesting attempts to construct
fully discrete Hamiltonian systems, but such ideas are still in the
embryonic stage of development.}, while the sequential nature of the
algorithm (changes are considered to only one spin at a time)
distinguishes it from the latter.  More pertinently, it should be noted
that the dynamics of the Metropolis algorithm do not necessarily have
any relation to physical dynamics -- other than the property of
converging to the equilibrium state.  The {\em physical\/} dynamical
properties of a system cannot be studied by (ordinary) Monte Carlo
methods; molecular dynamics, or some other such {\em microcanonical\/}
algorithm must be used.

Given the utility of the Ising model in studying the equilibrium
properties of materials, it should not be surprising that people have
tried very hard to extend such models to include real {\em dynamics}.
For a ferromagnet, this would mean being able to study the {\em
approach\/} of the magnetization to equilibrium.  One way to approach
this problem is to invent a {\em microcanonical\/} dynamics for the
Ising model.  Such a model would evolve the system in such a way as to
conserve energy globally and maintain the condition of {\em detailed
balance}.  If the probability that the system is in state $\bfS$ is
denoted by $P(\bfS)$, and if the probability of transition of the system
from state $\bfS$ to state $\bfS'$ is denoted by
$A(\bfS\rightarrow\bfS')$, then the latter condition, which ensures that
the dynamical process will converge to a Boltzmann-Gibbs equilibrium,
may be written
\[
P(\bfS)A(\bfS\rightarrow\bfS') = P(\bfS')A(\bfS'\rightarrow\bfS).
\]
Such a microcanonical dynamics for the Ising model was developed by
Creutz in 1983~\cite{bib:creutz}.  To localize the energy change
incurred by the flip of a spin, it was necessary to update the lattice
in a ``checkerboard'' pattern -- first updating the black squares and
then the red ones.  This has the effect of ``freezing'' a site's
neighbors while it is flipped, so that the energy change incurred by the
flip can be calculated in advance.  Because the algorithm is to be
microcanonical, this energy must be stored somewhere.  Creutz solved
this problem by allowing for each site to have a ``bank'' where it can
make local deposits and withdrawals of energy.  The bank must have a
finite capacity though, and if flipping the spin at a site would result
in an overflow or underflow of the local bank, the flip is not done;
otherwise, it is done.

Creutz demonstrated that his dynamics equilibrate the $D=2$ Ising model.
The lattice is initialized with a total energy (particles plus banks)
that never changes in the course of the simulation.  The portion of the
energy in the spins themselves, however, may change, and must therefore
be monitored by the simulation.  The important point, however, is that
Creutz' model is able to make a CA out of the Ising model.  The
checkerboard updating is not a real problem, since it can be
incorporated in the CA framework by including a ``time parity'' bit at
each site, initialized in a checkerboard pattern, and mandated to toggle
at each time step.  The value of this bit might then be used to
determine which sites will attempt to flip, and the rest is naught but
fully deterministic and reversible nearest-neighbor communication and
simultaneous updating at each site -- in other words, a reversible CA.

\subsection{From Ferromagnetism to Hydrodynamics}

The approach to equilibrium of a ferromagnet is certainly interesting,
but, because the order parameter (magnetization) is not a conserved
quantity, it happens relatively fast.  The path to equilibrium in
systems with conserved order parameters is generally more tortuous and
difficult -- and hence interesting -- because such systems are more
constrained.  Hence, for example, a viscous fluid at high Reynolds
number achieves equilibrium only after turbulent relaxation in which
intricate structures may be spawned across a wide range of scales in
length and time.

To be sure, it is possible to conserve the ferromagnetic order
parameter.  The so-called Kawasaki dynamics~\cite{bib:kawa} of the Ising
model flips only {\em pairs\/} of {\em oppositely directed\/} spins, so
that the total magnetization is conserved.  If this idea were combined
with Creutz' microcanonical dynamics, it would be possible to perform a
microcanonical simulation of a system with a conserved order parameter,
but I am not aware that this has been tried.  The more natural setting
for systems with conserved order parameters is hydrodynamics.  Fluids
generally have a conserved mass and momentum, and if compressibility
effects are considered, the conservation of energy will also play an
important role.

There was, however, a conceptual problem with the CA simulation of
fluids: Using lattice gases to model ferromagnets is intuitive enough
because it is not difficult to picture a regular array of magnetic
domains.  It seemed much less clear how to model a fluid on a lattice,
however, and almost a half a century elapsed between the development of
the Ising model and the first lattice-gas models of hydrodynamics.

\section{Hydrodynamic Lattice Gases}
\subsection{The Kadanoff-Swift Model}

The goal of a hydrodynamic lattice gas is to take the same
``minimalist'' approach to fluids that the Ising model takes to
ferromagnets.  The object is not a precise model of the dynamics at the
finest scales, but rather to invent a fictitious microdynamics whose
{\em coarse-grained\/} behavior -- in the thermodynamic limit -- lies in
the same universality class as the phenomenon under study.  Along these
lines, the approach that seems the most promising is to model the fluid
at the level of fictitious ``molecules'' that can move about and
collide, as they do in a real fluid, conserving mass, momentum and (for
compressible fluids) energy as they do so.

\begin{figure}
 \center{\mbox{\epsfxsize=3.0truein\epsffile{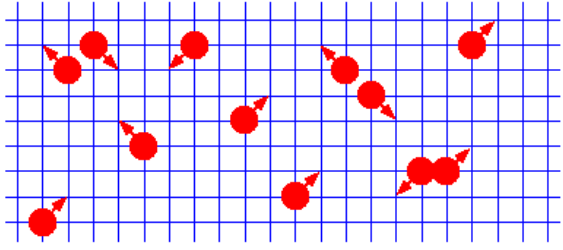}}}
  \caption{{\bf Kadanoff-Swift Model:} Particles carry a momentum vector
  oriented along diagonals.  They can stream in the direction of their
  momentum, diffuse without changing their momentum, or exchange their
  momentum with a neighboring particle.}
\label{fig:ks}
\end{figure}

The first attempt (known to me) along these lines was undertaken by
Kadanoff and Swift in 1968~\cite{bib:ks}.  They considered a $D=2$
Cartesian lattice, each site of which may be occupied by a particle, as
shown in Fig.~\ref{fig:ks}.  Each particle is tagged with one of four
momenta, oriented along the diagonals,
\begin{eqnarray*}
\bfc_1 &=& + \hat{\bfx} + \hat{\bfy} \\
\bfc_2 &=& - \hat{\bfx} + \hat{\bfy} \\
\bfc_3 &=& - \hat{\bfx} - \hat{\bfy} \\
\bfc_4 &=& + \hat{\bfx} - \hat{\bfy},
\end{eqnarray*}
where $\hat{\bfx}$ and $\hat{\bfy}$ are the unit vectors in the $x$ and
$y$ directions.  An Ising-like Hamiltonian is defined.  Then, at each
step, a particle is randomly selected and one of three things happen to
it:
\begin{itemize}
\item It moves in the direction of its velocity vector to the next site
where it could land without violating conservation of energy.
\item It diffuses, to any empty neighboring site, carrying its momentum
with it.
\item It exchanges momentum vectors with another neighboring particle.
\end{itemize}
The dynamics thus defined conserves mass, momentum and energy, and obeys
the principle of detailed balance.  Note that it is not technically a
cellular automaton, because of the sequential nature of the particle
updates, but it might be made into one by using some generalization of
checkerboard updating and/or Cruetz' ``banks.''  To my knowledge,
however, this has never been tried.

The Kadanoff-Swift (KS) model exhibits many features of real fluids,
such as sound-wave propagation, and long-time tails in velocity
autocorrelation functions.  As the authors noted, however, it does not
faithfully reproduce the correct equations of motion of a viscous (or,
for that matter, any other kind of) fluid.  In particular, the model
exhibits a strong lattice anisotropy; the decay of sound waves, for
example, depends on their direction of propagation with respect to the
underlying lattice.

\subsection{The HPP Model and Kinetic Theory}

The next advance in the lattice modelling of fluids came in the mid
1970's, when Hardy, de Pazzis and Pomeau introduced a new lattice model
with a number of innovations that warrant discussion here.  The HPP
model, named for its authors, also resides on a $D=2$ Cartesian lattice.
Particle velocities are taken from the set
\begin{eqnarray*}
\bfc_1 &=& +\hat{\bfx}\\
\bfc_2 &=& +\hat{\bfy}\\
\bfc_3 &=& -\hat{\bfx}\\
\bfc_4 &=& -\hat{\bfy},
\end{eqnarray*}
and there may be anywhere from zero to four particles at each site.  The
only restriction is that there may not be more than one particle with a
particular velocity, at a particular site, at a particular time.  This
``exclusion principle'' makes it possible to represent the state of any
site $\bfx$ by four bits; bit $n_j(\bfx,t)$, where $j\in\{1,2,3,4\}$,
encodes the presence ($1$) or absence ($0$) of a particle with velocity
$\bfc_j$ at site $\bfx$ and time $t$.

Given this representation, the dynamics is defined as follows: At each
time step, each site may experience a purely local {\em collision}, in
which its particles rearrange their velocity vectors in such a way as to
conserve mass and momentum.  A moment's thought indicates that the only
nontrivial way for this to happen is if exactly two particles enter a
site from opposite directions, and exit in the other two opposite
directions; for any other configuration, the particles simply retain
their incoming velocities, ``passing through'' each other without
interacting.  After the collisions, all particles ``stream'' to the site
in the direction of their velocity vector.  Note that all sites
experience collisions {\em simultaneously}, and all particles stream
simultaneously as well.  The HPP model can thus be regarded as a CA.

The above described dynamical rule can actually be expressed
algebraically as follows:
\begin{equation}
n_j(\bfx+\bfc_j, t+\Delta t) = n_j(\bfx,t) + \omega_j,
\label{eq:hppk}
\end{equation}
where $\Delta t$ is the time associated with one step of the dynamical
process (equal to unity if natural ``lattice units'' are adopted), and
where $\omega_j$ is called the {\em collision operator}.  If the
collision operator were not present, this equation would simply state
that a particle with velocity $\bfc_j$ will exist at site $\bfx+\bfc_j$
at time $t+\Delta t$ if it existed at site $\bfx$ one time step earlier.
This captures the streaming process.  To include the collisions,
$\omega_j$ must subtract (add) a particle from direction $j$ if the
incoming state will undergo a nontrivial collision that will deplete
(augment) that direction.  For the HPP rule described above, we have
\begin{eqnarray*}
\omega_1 &=& -n_1(1-n_2)n_3(1-n_4)+(1-n_1)n_2(1-n_3)n_4\\
\omega_2 &=& -n_2(1-n_3)n_4(1-n_1)+(1-n_2)n_3(1-n_4)n_1\\
\omega_3 &=& -n_3(1-n_4)n_1(1-n_2)+(1-n_3)n_4(1-n_1)n_2\\
\omega_4 &=& -n_4(1-n_1)n_2(1-n_3)+(1-n_4)n_1(1-n_2)n_3.
\end{eqnarray*}
Here we have made use of the fact that multiplication of bits is
equivalent to the logical ``and'' operation, substraction from one is
equivalent to the logical ``not'' operation, and addition of mutually
exclusive bits is equivalent to the logical ``or'' operation.  Thus, for
example, the quantity $(1-n_1)n_2(1-n_3)n_4$ is equal to one if
directions $2$ and $4$ are occupied, and $1$ and $3$ are not.  Thus,
this term appears with a plus sign in $\omega_1$ and $\omega_3$, since
the resulting collision augments those directions, and with a minus sign
in $\omega_2$ and $\omega_4$ since those directions are depleted.

This algebraic description of the exact microscopic motion of the
particles is somewhat akin to the Klimontovich description of continuum
kinetic theory~\cite{bib:klim}.  It is made somewhat simpler by the
discreteness of the spatial lattice, and the finite number of allowed
velocities.  To use this microscopic description to find the fluid
equations obeyed by the coarse-grained density and hydrodynamic velocity
requires all of the tricks of kinetic theory.  The first step is to
determine a kinetic equation for the single-particle distribution
function.  We define this by an {\em ensemble average}, supposing that
we have a large number of such lattices, with initial conditions sampled
from some (unspecified) distribution, and writing
\[
N_j(\bfx,t)\equiv\left\langle n_j(\bfx,t)\right\rangle,
\]
where the angle brackets denote the ensemble average.  Note that,
whereas $n_j$ is bit-valued, $N_j$ is real-valued.  By taking the
ensemble average of Eq.~(\ref{eq:hppk}), we arrive at
\begin{eqnarray*}
N_j(\bfx+\bfc_j, t+\Delta t)
&=&
 N_j(\bfx,t)
 +\left\langle\omega_j\right\rangle\\
&=&
 N_j(\bfx,t)
 -\left\langle
   n_j(1-n_{j+1})n_{j+2}(1-n_{j+3})
  \right\rangle\\
& &
 +\left\langle
   (1-n_j)n_{j+1}(1-n_{j+2})n_{j+3}
  \right\rangle ,
\end{eqnarray*}
where all subscripts are evaluated modulo $4$.  At this point, we see
that we have a problem.  The collision operator is nonlinear in the
$n_j$'s, and the average of a product is not generally equal to the
product of the averages -- not if the quantities involved are
correlated.  Thus, the dynamical equation of the $N_j$'s will involve
averages of products, such as $\left\langle n_j n_{j+2}\right\rangle$.
To know these, it would be necessary to write kinetic equations for
these two-point correlations, but these will involve still higher
correlations, etc.  This infinite series of equations is the lattice-gas
analog of the BBGKY hierarchy of kinetic theory.

To truncate this hierarchy and obtain a closed equation for the $N_j$'s,
it is necessary to make a physical approximation: We shall assume that
the particles entering a collision are uncorrelated.  This approximation
is tantamount to Boltzmann's famous {\em molecular chaos\/} assumption.
It is unlikely to be true, especially for high densities and
low-dimensional lattices.  For now we note that, under this assumption,
it is possible to replace the average of products above by the product
of averages, resulting in a closed equation for the single-particle
distribution function,
\begin{eqnarray}
N_j(\bfx+\bfc_j, t+\Delta t)
&=&
 N_j(\bfx,t)
 +\Omega_j\nonumber\\
&=&
 N_j(\bfx,t)
 - N_j(1-N_{j+1})N_{j+2}(1-N_{j+3})\\
& &
 + (1-N_j)N_{j+1}(1-N_{j+2})N_{j+3}.
\label{eq:lboltz}
\end{eqnarray}
This is called the {\em lattice-Boltzmann equation}.  From this, it is
possible, using the Chapman-Enskog analysis of classical kinetic theory,
to derive the hydrodynamic equations obeyed by the mass density,
\begin{equation}
\rho(\bfx,t)\equiv\sum_i N_i(\bfx,t)
\label{eq:dens}
\end{equation}
and the momentum density,
\begin{equation}
\rho(\bfx,t)\bfu(\bfx,t)\equiv\sum_i \bfc_i N_i(\bfx,t).
\label{eq:hvel}
\end{equation}
Note that Eqs.~(\ref{eq:dens}) and (\ref{eq:hvel}) are the
discrete-velocity analog of the usual integration over velocity space to
obtain the hydrodynamic densities.  In this way, the fully compressible
hydrodynamic equations for the HPP lattice gas were worked out in the
original papers by HPP~\cite{bib:hpp}.  For our purposes, we note that
the result of this exercise in the {\em incompressible\/} limit is
$\bfnabla\cdot\bfu=0$ and
\begin{eqnarray}
\frac{\partial u_x}{\partial t} +
g(\rho)\frac{\partial}{\partial x} u_y^2 &=&
-\frac{1}{\rho}\frac{\partial P}{\partial x} +
\nu(\rho)\frac{\partial^2 u_x}{\partial x^2} -
\left(\nu(\rho)+\frac{1}{2}\right)
\frac{\partial^2 u_y}{\partial x\partial y}\nonumber\\
\frac{\partial u_y}{\partial t} +
g(\rho)\frac{\partial}{\partial y} u_x^2 &=&
-\frac{1}{\rho}\frac{\partial P}{\partial y} +
\nu(\rho)\frac{\partial^2 u_y}{\partial y^2} -
\left(\nu(\rho)+\frac{1}{2}\right)
\frac{\partial^2 u_x}{\partial x\partial y},
\label{eq:hpphyd}
\end{eqnarray}
where $P$ is the pressure, and we have defined the functions
\[
g(\rho)\equiv\frac{1-\rho/2}{1-\rho/4}
\]
and
\[
\nu(\rho)\equiv\frac{1}{2\rho (1-\rho/4)}-\frac{1}{2}.
\]

Eqs.~(\ref{eq:hpphyd}) has some superficial resemblence to the
Navier-Stokes equations of viscous fluid dynamics, but closer inspection
reveals some important differences.  Like the KS model studied earlier,
the HPP model gives rise to anisotropic hydrodynamic equations that are
not invariant under a global spatial rotation.  They involve derivatives
with respect to $x$ and $y$ in combinations that cannot be expressed in
terms of the familiar $\bfnabla$ operator of vector calculus.  Rather,
the grid coordinates $x$ and $y$ have a preferred status.  Hence, for
example, the drag of a KS or HPP fluid as it flows past a generic
obstacle will depend on that obstacle's angle of orientation with
respect to the underlying lattice.  At the time, this was not considered
a problem, since the real purpose of the KS and HPP models was to study
the statistical physics of fluids, and both models could do this well.
Traditional computational fluid dynamicists, however, were not inclined
to take notice of this as a serious numerical method unless and until a
way was found to remove the unphysical anisotropy.

\subsection{The FHP Model}

\begin{figure}
  \centering
  \begin{picture}(300,160)(0,0)
    \put(75.,150.){\makebox(0,0){In}}
    \put(225.,150.){\makebox(0,0){Out}}
    \put(225.,90.){\makebox(0,0){\footnotesize or}}
    \put(180.,90.){
      \begin{picture}(50,50)(0,0)
        \put(0.,0.){\vector(2,3){12.5}}
        \put(0.,0.){\vector(-2,-3){12.5}}
        \put(0.,0.){\circle*{5.}}
      \end{picture}}
    \put(270.,90.){
      \begin{picture}(50,50)(0,0)
        \put(0.,0.){\vector(-2,3){12.5}}
        \put(0.,0.){\vector(2,-3){12.5}}
        \put(0.,0.){\circle*{5.}}
      \end{picture}}
    \put(225.,30.){
      \begin{picture}(50,50)(0,0)
        \put(0.,0.){\vector(1,0){25.}}
        \put(0.,0.){\vector(-2,3){12.5}}
        \put(0.,0.){\vector(-2,-3){12.5}}
        \put(0.,0.){\circle*{5.}}
      \end{picture}}
    \put(75.,90.){
      \begin{picture}(50,50)(0,0)
        \put(25.,0.){\vector(-1,0){22.5}}
        \put(-25.,0.){\vector(1,0){22.5}}
        \put(0.,0.){\circle*{5.}}
      \end{picture}}
    \put(75.,30.){
      \begin{picture}(50,50)(0,0)
        \put(25.,0.){\vector(-1,0){22.5}}
        \put(-12.5,21.65){\vector(2,-3){11.1132}}
        \put(-12.5,-21.65){\vector(2,3){11.1132}}
        \put(0.,0.){\circle*{5.}}
      \end{picture}}
  \end{picture}
  \caption{{\bf Nontrivial collisions of the FHP lattice gas:}  The
    dynamics takes place on a triangular lattice in two dimensions.  The
    symmetric three-body collision is necessary to avoid a third spuriously
    conserved component of momentum.  Asymmetric three-body and four-body
    collisions are also possible, but not illustrated here.}
  \label{fig:fhp}
\end{figure}
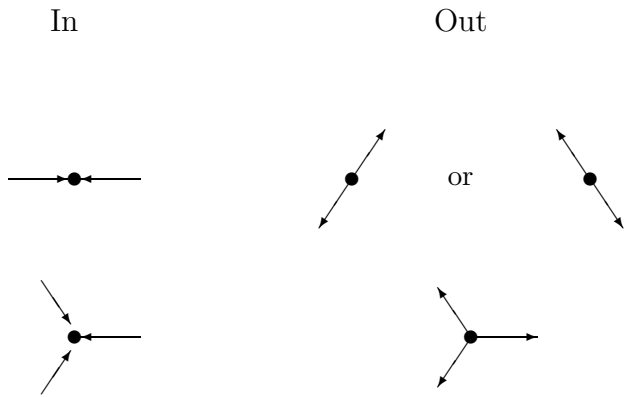

Another thirteen years passed from the introduction of the HPP model to
the solution of the anisotropy problem in 1986 by Frisch, Hasslacher and
Pomeau~\cite{bib:fhp}, and simultaneously by Wolfram~\cite{bib:swolf}.
The FHP lattice gas, named after the authors of the first reference
given above, is very similar to that of HPP in that the evolution
proceeds by alternating collision and streaming steps -- hence, it is
again a cellular automaton.  The only real difference is that it is
based on a triangular lattice instead of a Cartesian one, as shown in
Fig.~\ref{fig:fhp}.  Now one would expect that a six-fold symmetric
lattice would give rise to a more isotropic model than a four-fold
symmetric one.  The surprising result of the 1986 studies, however, is
that the six-fold version does not only improve the isotropy -- it
yields {\em perfect} isotropy!

To see why isotropy is recovered on a triangular lattice, we generalize
the results of the Chapman-Enskog analysis mentioned above.  It turns
out that the most general form for the viscous term in the hydrodynamic
equations above is
\[
\frac{\partial u_i}{\partial t}+\cdots =
c_{ijkl}\nabla_j\nabla_k u_l,
\]
where $c_{ijkl}$ is a fourth-rank tensor that is constructed from the
lattice vectors, and hence shares all of their symmetries.  If the
lattice is invariant under rotation by $60^\circ$ or $90^\circ$, so then
will be the components of this tensor.  Note that if this tensor is
isotropic -- so that it is expressible in terms of the Kronecker delta
-- then we have most generally
\[
c_{ijkl} = \nu\delta_{jk}\delta_{il}+
           \mu\delta_{ij}\delta_{kl}+
           \mu\delta_{ik}\delta_{jl},
\]
whence our hydrodynamic equation becomes
\[
\frac{\partial\bfu}{\partial t}+\cdots =
\nu\nabla^2\bfu + 2\mu\bfnabla\left(\bfnabla\cdot\bfu\right).
\]
This form, expressible with vector notation, is indeed isotropic and the
two terms on the right may be identified with the shear and bulk
viscosity terms of the Navier-Stokes equation, respectively.  Thus, a
sufficient condition for the attainment of isotropy is to ensure that
the $c_{ijkl}$ tensor is isotropic.  It turns out that the only
rank-four tensors in two dimensions that are invariant under $60^\circ$
rotation are isotropic; by contrast, there exist rank-four tensors in
two dimensions that are invariant under $90^\circ$ rotation but which
are not isotropic (cannot be expressed in terms of the Kronecker delta).
Thus, generically, isotropy will be recovered on a triangular grid and
not on a Cartesian one.

In the incompressible limit, the result of the Chapman-Enskog analysis
for the FHP fluid is then $\bfnabla\cdot\bfu=0$ and
\begin{equation}
\frac{\partial\bfu}{\partial t} + g(\rho)\bfu\cdot\bfnabla\bfu =
-\bfnabla P + \nu(\rho)\nabla^2\bfu,
\label{eq:ns}
\end{equation}
where $g(\rho)$ is defined as it was for the HPP model, and $\nu(\rho)$
is a well-behaved function of density whose precise form depends on the
particulars of the collision rules (e.g., whether or not three- or
four-body collisions are considered).  Thus, isotropy is recovered, and
there is no problem writing Eq.~(\ref{eq:ns}) in the usual notation of
vector calculus, as we have done.

There is, however, one last lingering problem: The factor of $g(\rho)$
in front of the inertial term (second term on the left-hand side) is
equal to unity in the real Navier-Stokes equations.  This is a
consequence of Galilean invariance (GI) which, like isotropy, is yet
another Lie-group symmetry possessed by the real Navier-Stokes
equations, but which we may not take for granted in our lattice model.
To see that the convective derivative operator
\begin{equation}
\frac{D}{Dt}\equiv
\frac{\partial}{\partial t} + \bfu\cdot\bfnabla
\label{eq:cd}
\end{equation}
has GI, we consider the transformation of coordinates
\begin{eqnarray*}
\bfx' &=& \bfx - \bfV t\\
t' &=& t,
\end{eqnarray*}
where $\bfV$ is a constant vector.  This has inverse
\begin{eqnarray*}
\bfx &=& \bfx' + \bfV t'\\
t &=& t'.
\end{eqnarray*}
Then the derivatives transform as follows:
\begin{eqnarray*}
\frac{\partial}{\partial t}
&=&
\frac{\partial t'}{\partial t}
\frac{\partial}{\partial t'} +
\frac{\partial\bfx'}{\partial t}
\cdot\bfnabla'
=
\frac{\partial}{\partial t'} - \bfV\cdot\bfnabla'\\
\bfnabla
&=&
\left(\bfnabla t'\right)
\frac{\partial}{\partial t'} +
\left(\bfnabla\bfx'\right)\cdot\bfnabla' =
\bfnabla',
\end{eqnarray*}
and the hydrodynamic velocity transforms as follows
\[
\bfu =
\frac{d\bfx}{dt} =
\frac{d}{dt'}\left(\bfx' + \bfV t'\right) =
\frac{d\bfx'}{dt'} + \bfV =
\bfu' + \bfV.
\]
Under this Galilean transformation, we see that
\[
\frac{\partial}{\partial t} + \bfu\cdot\bfnabla =
\frac{\partial}{\partial t'} - \bfV\cdot\bfnabla' +
\left(\bfu' + \bfV \right)\cdot\bfnabla' =
\frac{\partial}{\partial t'} + \bfu'\cdot\bfnabla',
\]
so the form of the convective derivative operator is indeed Galilean
invariant.  The presence of the $g(\rho)$ factor invalidates this
argument.  The lack of GI in the FHP fluid is due to the fact that the
lattice itself constitutes a preferred reference frame.

We can now see the isotropy and GI problems in a unified light:
Lie-group symmetries are often responsible for particular features of
hydrodynamic equations; for example, isotropy implies that the equation
can be written in the notation of vector calculus, and GI implies that
the convective derivative will have the form of Eq.~(\ref{eq:cd}).  Such
symmetries may be broken by the presence of the fixed lattice, and when
this happens the corresponding features of the hydrodynamic equations
may be destroyed.  We managed to recover isotropy by noting that the
maximum rank of the tensors in our hydrodynamic equation is four, and
that rank-four tensors with $60^\circ$ rotational invariance are always
isotropic.  How do we fix the GI problem?  There are several ways to do
so.  For incompressible flow, for which $\rho$ and hence $g(\rho)$ are
constant, the easiest solution is to scale $\bfu$ and the pressure $P$
by $g(\rho)$; since the term in which $g(\rho)$ appears is the only one
that is quadratic in $\bfu$, this works handily.  Another approach
introduces additional particle velocities to force $g(\rho)$ to unity.

Another consideration is the functional form of the shear viscosity,
$\nu(\rho)$.  For lattice gases that satisfy detailed balance, this
quantity will have a minimum as a function of $\rho$.  This minimum
value of the viscosity sets the maximum value of the Reynolds number
that may be simulated on a given size lattice.

The result of this 1986 work was a cellular automaton model whose
coarse-grained behavior was that of a Navier-Stokes fluid.  Now the
computational fluid dynamicists took notice, and there was a
five-year-long flurry of activity in the field.  Some of the
accomplishments during this time were
\begin{enumerate}
\item \label{it:td} the extension of the model to three dimensions.
\item simulations of flow past various obstacles, and comparisons with
other more conventional methods of computational fluid dynamics (CFD).
\item clever variations on the collision rules of the basic FHP model
intended to achieve lower viscosity minima, and experimental tests of
the functional form of $\nu(\rho)$.
\item careful measurements of long-time tails in velocity
autocorrelation functions, and finite-size corrections to the viscosity.
\item \label{it:cf} extensions of the model to simulate complex fluid
hydrodynamics, including interfaces and surface tension.
\item a host of algorithmic tricks, and even special-purpose hardware,
to simulate such hydrodynamic lattice gases on parallel computers.
\end{enumerate}
The number of papers produced in this period is far too great to review
here; the interested reader is referred to the excellent text by Rothman
and Zaleski~\cite{bib:rz} and the secondary references therein, for a
review of this work.  Items \ref{it:td} and \ref{it:cf} on the above
list will be discussed briefly below in Subsection~\ref{ssec:fchc} and
Section~\ref{sec:cf}, respectively.

One reason that hydrodynamic lattice gases captured peoples' imagination
at this time was that they represented an altogether new way of doing
CFD -- and, indeed, computational physics in general.  Conventional CFD
methods {\em began} with the Navier-Stokes equations, and discretized
them in one of a variety of ways.  Lattice gases, by contrast, defined a
kind of particle kinetics from which the Navier-Stokes equations were
{\em emergent} -- just as they are emergent for a real fluid.  There are
definite advantages to such a ``physical'' approach, aside from its
undeniable aesthetic appeal.  For example, one often overlooked
advantage of lattice-gas models is their unconditional stability.  The
Navier-Stokes equations have a basis in kinetic theory, as the behavior
of a system of particles whose collisions conserve mass and momentum.
The fact that these underlying collisions obey a detailed-balance
condition ensures the validity of the $H$ theorem, the
fluctuation-dissipation theorem, Onsager reciprocity, and a host of
other critically important properties with macroscopic consequences.
When its kinetic origins are cavalierly ignored, and the Navier-Stokes
equations are ``chopped up'' into finite-difference schemes, these
important properties can be lost.  The discretized evolution equations
need no longer satisfy an $H$ theorem, and the notion of underlying
fluctuations may lose meaning altogether.  As the first practioneers of
finite-difference simulations on digital computers found in the 1940's
and 1950's, the result can be the development and growth of
high-wavenumber {\em numerical instabilities}, and indeed these have
plagued essentially all CFD methodologies in all of the decades since.
Such instabilities are entirely unphysical because they represent a
clear violation of the $H$ theorem; indeed, the Second Law of
Thermodynamics would preclude their occurrence.  Numerical analysts have
responded to this problem with textbooks full of ways to ``patch up''
these anomalies -- including upwind differencing, artificial viscosity,
and a host of other very clever tricks -- but from a {\em physicist's\/}
point of view it would have been much better if the original
discretization process had retained more of the underlying {\em
physics}, so that these problems had not occurred in the first place.
Lattice gases represented an important first step in this direction.  As
was shown shortly after their first applications to hydrodynamics, they
can be constructed with an $H$ theorem that rigorously precludes any
kind of numerical instability.  More glibly stated, lattice gases avoid
numerical instabilities in precisely the same way that Nature herself
does so.

Even the computer implementation of a hydrodynamic lattice gas was
novel.  All other CFD methodologies make use of floating-point numbers
to represent real quantities.  In such floating-point numbers, some bits
represent the mantissas, others the exponents and others the signs.  A
lattice gas, by contrast, can use a representation of one bit per
particle.  All bits thus play an equal role in some sense; this idea is
sometimes colorfully referred to as ``bit democracy.''

In spite of these successes, the use of lattice gases for the simulation
of simple (single-phase) Navier-Stokes fluids declined substantially in
the early 1990's.  Ironically, it was largely supplanted by the direct
floating-point computer simulation of the lattice Boltzmann equation,
Eq.~(\ref{eq:lboltz}).  The lattice Boltzmann equation also has kinetic
underpinnings in a sense, though its representation is that of a
single-particle distribution function, rather than a full particle-level
description.  Most importantly, the lattice Boltzmann framework allows
for greater accuracy than LGA by effectively eliminating kinetic
fluctuations.  Measured quantities are therefore much less noisy, and
require less ensemble averaging to compute accurately.  For certain
complex fluids such as microemulsions, however, lattice gases remain a
very effective simulation methodology.

\subsection{The FCHC Model}
\label{ssec:fchc}

To find a lattice-gas model of isotropic three dimensional
hydrodynamics, it was necessary to find a $D=3$ lattice, under whose
symmetry group the only {\em isotropic\/} rank-four tensors are
invariant.  The trouble is that no such regular lattice exists in three
dimensions.  In 1987, however, Frisch et al.~\cite{bib:fchc} noticed
that a lattice with the required symmetry does exist in four dimensions.
It is called the face-centered hypercubic (FCHC) lattice, and is
self-dual with 24 lattice vectors per site.  The 24 lattice vectors can
be arranged in three groups of eight, such that any two groups of eight
comprise the 16 vertices of a regular four-dimensional hypercube, and
the vectors of the third group of eight point in the direction of the
centers of the faces of that hypercube (hence, the name FCHC).

The 24 lattice vectors are most easily enumerated by taking all integer
quadruples $(i,j,k,l)$ that lie at a distance $\sqrt{2}$ from the
origin.  Clearly, two of the integers components must be zero, and the
other two must be $\pm 1$.  There are $\mychoose{4}{2}=6$ ways to choose
which two are zero, and $2^2=4$ ways of assigning $\pm 1$ to the other
two, for a total of 24.

It was noticed~\cite{bib:fchc} that by projecting this lattice back to
three dimensions -- by, say, ignoring the fourth coordinate -- a simple
set of $D=3$ lattice vectors was obtained which worked.  The three
dimensional lattice thus obtained is not a regular lattice, in that not
all the lattice vectors have the same length, and some of them have
multiplicity two (because they correspond to two {\em different\/}
vectors on the four-dimensional lattice), but it works nevertheless.
The projected lattice vectors are thus
\begin{center}
\begin{tabular}{r|rrr}
$j$ & $c_{jx}$ & $c_{jy}$ & $c_{jz}$ \\
\hline
0,1   &+1 & 0 & 0 \\
2,3   &-1 & 0 & 0 \\
4     & 0 &+1 &+1 \\
5     & 0 &+1 &-1 \\
6     & 0 &-1 &+1 \\
7     & 0 &-1 &-1 \\
\hline
8,9   & 0 &+1 & 0 \\
10,11 & 0 &-1 & 0 \\
12    &+1 & 0 &+1 \\
13    &+1 & 0 &-1 \\
14    &-1 & 0 &+1 \\
15    &-1 & 0 &-1 \\
\hline
16,17 & 0 & 0 &+1 \\
18,19 & 0 & 0 &-1 \\
20    &+1 &+1 & 0 \\
21    &+1 &-1 & 0 \\
22    &-1 &+1 & 0 \\
23    &-1 &-1 & 0 \\
\end{tabular}
\end{center}
Note that four of them have multiplicity two.  Also note that we have
listed them in the three groups of eight, described above.  A more
geometric description of the FCHC lattice and its use for lattice-gas
simulations is given in the reference by Adler et al.~\cite{bib:adler}.

\section{Applications to Complex Fluids}
\label{sec:cf}

The simulation of the hydrodynamics of {\em complex fluids}, such as
immiscible flow, coexisting phases, emulsions, colloids, liquid
crystals, gels and foams is one of the principal outstanding challenges
of computational condensed matter physics.  Hydrodynamic equations for
such materials are often not known or ill-posed, so that
finite-difference discretizations are not even an option.  Molecular
dynamics can, of course, be employed, but it advances in time steps that
are typically between $10^{-2}$ and $10^{-3}$ of a mean-free path.  In
this context, lattice gases offer the possibility of an {\em
inexpensive\/} molecular dynamics -- one for which particles are still
discrete, but can advance in steps on the order of a mean-free time
(since the particles typically suffer a collision at each time step).
Earlier in this review, we mentioned that lattice models can be useful
for studying the equilibrium properties of complex fluids, and we
considered the model of Widom in some detail.  We finish this survey by
demonstrating how lattice gas hydrodynamics have begun to explore the
{\it hydrodynamics} of such materials.

In 1988, Rothman and Keller~\cite{bib:rk} introduced a lattice-gas model
of immiscible flow, such as that of oil and water.  They accomplished
this by tagging the lattice gas particles with two ``colors,'' to
distinguish oil and water.  Their collisions were required to conserve
the total mass of each color separately, as well as the total momentum.
They then skewed the collision outcomes to favor those that send
particles towards sites dominated by other particles of the same color.
This affinity of particles for other particles of the same color gives
the two phases cohesion, and the interface surface tension.  Let us call
the two colors ``red'' and ``blue,'' and use $n^R_i(\bfx,t)$ and
$n^B_i(\bfx,t)$, respectively, to denote the occupation number of each
color in velocity direction $i$, at site $\bfx$ and at time $t$.  Then
Rothman and Keller defined a {\em color field}
\[
\bfE\equiv\sum_i\bfc_i\left[\sum_j\left(n^R_j-n^B_j\right)\right]
\]
at each site that points in the direction of increasing (red minus blue)
color, and a {\em color flux}
\[
\bfJ\equiv\sum_i\bfc_i\left(n^R_i-n^B_i\right)
\]
for each possible {\em outgoing\/} state.  They then chose the outcome
that minimized the {\em color work},
\begin{equation}
H = -\bfJ\cdot\bfE.
\label{eq:colwork}
\end{equation}
More generally, Chen et al.~\cite{bib:chen} pointed out that one should
assign probabilities to each of the possible outcomes based on the
Boltzmann weights $\exp\left(-\frac{H}{k_BT}\right)$, where $T$ is a
temperature.  This model has been studied extensively
(see~\cite{bib:rz}), and it exhibits phase separation and surface
tension.

Another rather different model of two coexisting phases, such as water
and water vapor, was worked out by Appert and Zaleski in
1990~\cite{bib:az}.  The model gives particles an attraction by allowing
mass- and momentum-conserving collisions between certain configurations
of particles at different sites.  This model has also been extensively
studied (again, see~\cite{bib:rz}).  Both this model and that of Rothman
and Keller have been cast in the lattice Boltzmann framework as
well~\cite{bib:shanchen,bib:yeomans}; indeed, this remains a very active
area of study.

Neither the Rothman-Keller and Appert-Zaleski models, as originally
posed, obeyed the principle of detailed balance.  Since this is
important, for all of the reasons described above, some effort has been
devoted to restoring detailed balance to hydrodynamic lattice-gas models
with interaction between particles at different sites.  This turns out
not to be easy.  The best attempts to date have taken as a starting
point a 1988 method due to Colvin et al.~\cite{bib:mdsq}, called {\sl
Maximally Discretized Molecular Dynamics} or ``$(MD)^2$.''  This is
essentially a return to the Kadanoff-Swift representation with a maximum
of one particle per site, and {\sl sequential\/} propagation of
particles.  (Thus, this model is no longer, strictly speaking, a
cellular automaton.)  Like the KS model, particles are allowed to
exchange momentum with their neighbors, and they can step in the
direction of their momentum to the nearest empty site.  A hard-sphere
interaction is introduced that may extend over more than one site.
Mass, momentum and energy are exactly conserved, and detailed balance is
maintained.

Arbitrary interaction potentials -- possibly beyond just the nearest
neighbor in range -- were added to the $(MD)^2$ model by Gunn et
al.~\cite{bib:gunn} in 1993, and this model remains near the state of
the art~\footnote{The algorithm in the paper by Gunn et al. used a
continuum-valued velocity, but this is not an essential feature.}.  The
collisions in this model do not change the energy, and may therefore be
unconditionally accepted.  The propagation step, however, may change the
energy, and the algorithm can be implemented in either a microcanonical
version (in which the propagation is accepted only if the energy change
is zero), or a canonical version (in which it is accepted according to
the Metropolis algorithm).

Finally, we note that many variants of these algorithms exist and are
useful for various purposes.  Boghosian et al.~\cite{bib:bce} developed
a variation of the Rothman-Keller model that allows for the inclusion of
a surfactant phase, allowing the simulation of microemulsions.  To see
how this works, first note that the quantity
\[
\sum_i \left(n^R_i-n^B_i\right)
\]
can be thought of as a {\em color charge\/} in the Rothman-Keller model,
in that its current is the color flux, and its vector-weighted sum over
neighbors is the color field.  In this context, surfactant particles are
introduced as {\sl color dipoles\/}, and numerous terms are added to the
Hamiltonian, Eq.~(\ref{eq:colwork}), to account for the color-dipole
interaction that makes the surfactant prefer to live on the interface,
and the dipole-dipole interaction that gives rise to the curvature
energy.  Details are given in the reference~\cite{bib:bce}.

\begin{figure}
 \center{\mbox{\epsfxsize=3.0truein\epsffile{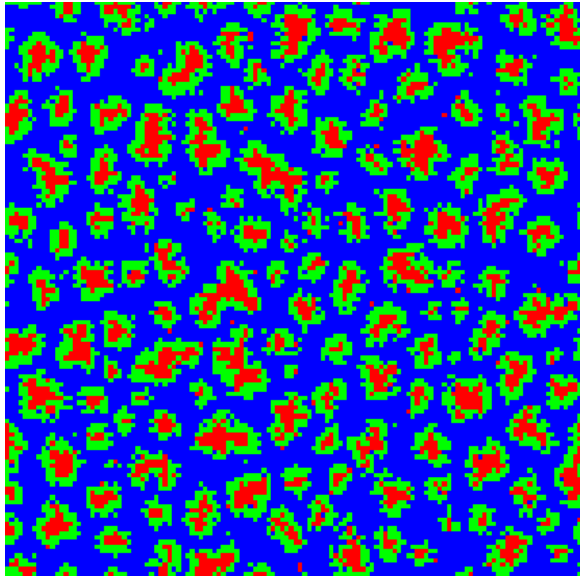}}}
 \caption{{\bf Growth of emulsion droplets} in three-component mixture
 of water, oil and amphiphile in two dimensions.  Lattice gases have
 been used to study the growth and saturation of emulsion droplets.}
\label{fig:sep}
\end{figure}

\begin{figure}
 \center{\mbox{\epsfxsize=3.0truein\epsffile{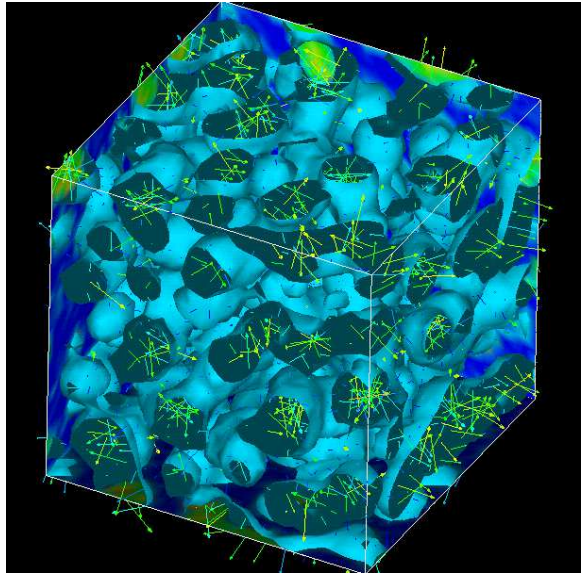}}}
 \caption{{\bf Growth of wormlike micelle phase} in binary mixture of
 water and amphiphile in three dimensions.  The arrows denote the
 direction of orientation of individual amphiphile particles.  This
 structure self-organized from randomly homogenized initial conditions.}
\label{fig:worm}
\end{figure}

\begin{figure}
 \center{\mbox{\epsfxsize=3.0truein\epsffile{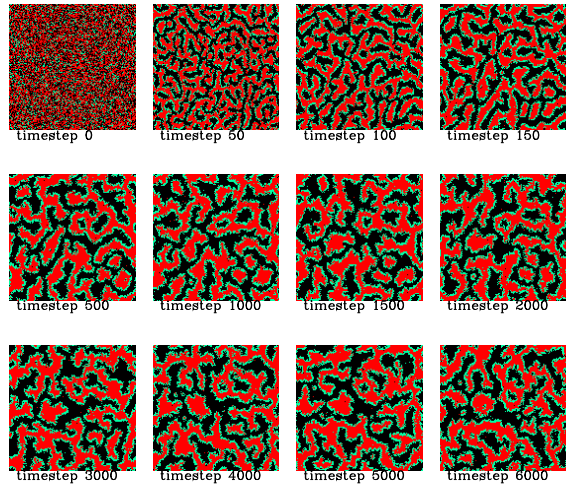}}}
 \caption{{\bf Growth of sponge phase} in three-component mixture of
 water, oil and amphiphile in two dimensions.  Again, this structure
 self-organized from random initial conditions.}
\label{fig:spo}
\end{figure}

\begin{figure}
 \center{\mbox{\epsfxsize=3.0truein\epsffile{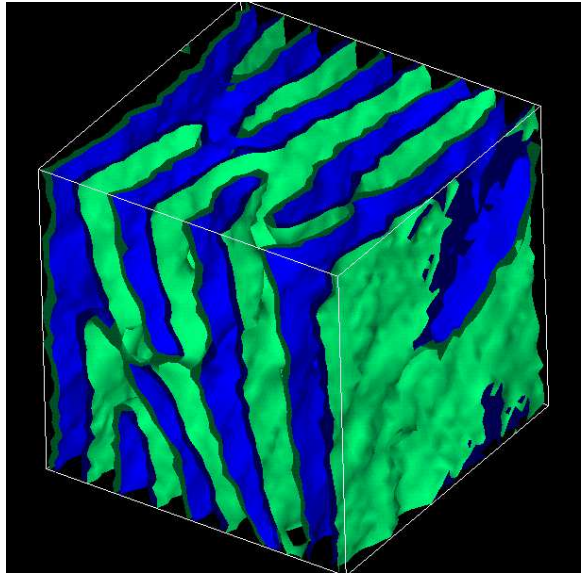}}}
 \caption{{\bf Growth of lamellar phase} in three-component mixture in
 three dimensions.  Only the interface is rendered for clarity.}
\label{fig:lam}
\end{figure}

\begin{figure}
 \center{\mbox{\epsfxsize=3.0truein\epsffile[50 350 515 750]{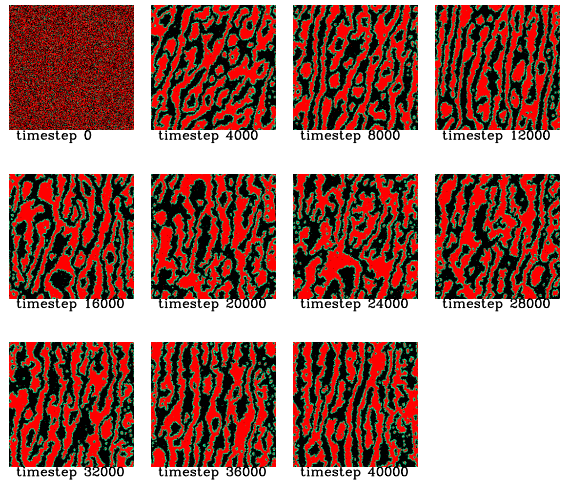}}}
 \caption{{\bf Shear-induced sponge-to-lamellar phase transition} in a
 three-component mixture in two dimensions, with an applied shear.}
\label{fig:stl}
\end{figure}

The $D=2$ model has been studied in some detail, and preliminary $D=3$
results have been obtained as of this writing.  The model is able to
track the formation and saturation of droplet (Fig.~\ref{fig:sep},
$D=2$), wormlike-micelle (Fig.~\ref{fig:worm}, $D=3$), sponge
(Fig.~\ref{fig:spo}, $D=2$) and lamellar (Fig.~\ref{fig:lam}, $D=3$)
phases, and the time dependence of this saturation has been
studied~\cite{bib:sep}.  Interfacial fluctuations in the presence of
surfactant have been studied with this model~\cite{bib:ifl}, and it has
also been used for the first simulations of the shear-induced
sponge-to-lamellar phase transition~\cite{bib:stl} (Fig.~\ref{fig:stl},
$D=2$).  Note that all of these applications involve nonequilibrium or
dynamical processes that have previously been difficult to address.  The
most lengthy MD simulations to date are barely able to see the
self-assembly of a single emulsion droplet.  The lattice-gas method, by
contrast, is able to study the growth and saturation of many such
droplets and larger structures, as can be seen in these figures.

\section{Conclusion}

We have reviewed the use of lattice gases for physical problems, with
emphasis on their applications to hydrodynamics.  We have traced the
evolution of hydrodynamic lattice gases in the 1960's and 1970's,
leading up to the development of the FHP model in 1986.  We have then
seen some of the attempts to add interactions -- possibly between
particles at different sites -- to such models, in order to simulate
complex-fluid hydrodynamics; we have also discussed the loss of detailed
balance in such models, and its recovery by the $(MD)^2$ algorithm and
its variants.  Though much of the CFD-related activity in this field has
migrated to lattice Boltzmann methods, it is this author's belief that
lattice gases remain very well suited for the role of ``inexpensive
molecular dynamics'' for the hydrodynamic simulation of complex fluids.

In addition, it seems likely that this kind of simulation can still
benefit from the development of special-purpose hardware.  Interest in
this area has been somewhat stifled to date because the ongoing meteoric
rise in workstation performance (per unit cost) has made it dangerous to
try to develop any kind of hardware outside of the mainstream.  Sooner
or later, however, workstation performance will begin to saturate, and
the very real differences between hardware optimized for lattice gases
and for general-purpose floating-point calculations will once again
become a target on which hardware designers might well focus some
activity.

\section*{Acknowledgements}

The author would like to thank Raissa d'Souza, Norman Margolus and Peter
Coveney for their review of the manuscript and helpful comments.  The
three-dimensional images were made by Andrew Emerton, Peter Love and
David Bailey of Oxford University.


\begin{thebibliography}{999}
  \parindent=.6em 
\bibitem{bib:lenz} W. Lenz,
    {\em Phys. Zeitschrift} {\bf 21} (1920) 613.
\bibitem{bib:ising} E. Ising,
    {\em Z. der Physik}  {\bf 31} (1925) 253.
\bibitem{bib:onsager} L. Onsager,
    {\em Phys. Rev.} {\bf 65} (1944) 117.
\bibitem{bib:zama}  A.B. Zamalodchikov, ``Integrals of Motion and
    $S$-Matrix of the (Scaled) $T=T_c$ Ising Model with Magnetic
    Field,'' {\em Int. J. Mod. Phys. A} {\bf 4} (1989) 4235.
\bibitem{bib:widom} B. Widom, ``Lattice Model of Microemulsions,''
    {\em J. Chem. Phys.} {\bf 84} (1986) 6943--6954.
\bibitem{bib:mmc} N. Metropolis, A. Rosenbluth,
    M. Rosenbluth, A. Teller, and E. Teller, ``Equation of State
    Calculations by Fast Computing Machines,'' {\em J. Chem Phys.} {\bf
    21} (1953) 1087-1092.
\bibitem{bib:creutz} M. Creutz, ``Microcanonical Monte Carlo Simulation,''
    {\em Phys. Rev. Lett.} {\bf 50} (1983) 1411--1414; ``Deterministic
    Ising Dynamics,'' {\em Annals of Physics} {\bf 167} (1986) 62--72.
\bibitem{bib:kawa} Kawasaki, K.,
    {\em Ann. Phys.} (N.Y.) {\bf 61} (1970) 1.
\bibitem{bib:ks} L.P. Kadanoff and J. Swift,
    {\em Phys. Rev.} {\bf 165} (1968) 310.
\bibitem{bib:klim} Yu. L. Klimontovich,
    ``The Statistical Theory of Nonequilibrium Processes in a Plasma,''
    Chapter 4, M.I.T. Press, Cambridge, Massachusetts (1967).
\bibitem{bib:hpp} J. Hardy, O. de Pazzis and Y. Pomeau, ``Time Evolution
    of a Two-Dimensional Model System. I. Invariant States and Time
    Correlation Functions,'' {\em J. Math. Phys.} {\bf 14} (1973)
    1746-1759; ``Molecular Dynamics of a Classical Lattice Gas:
    Transport Properties and Time Correlation Functions,'' {\em
    Phys. Rev. A} {\bf 13} (1976) 1949.
\bibitem{bib:fhp} U. Frisch, B. Hasslacher, and Y. Pomeau, ``Lattice-Gas
    Automata for the Navier-Stokes Equation,'' {\em Phys. Rev. Lett.}
    {\bf 56} (1986) 1505.
\bibitem{bib:swolf} S. Wolfram, ``Cellular Automaton Fluids I. Basic
    Theory,'' {\em J. Stat. Phys.} {\bf 45} (1986) 471.
\bibitem{bib:rz} D.H. Rothman and S. Zaleski,
    ``Lattice-Gas Automata: Simple Models of Complex Hydrodynamics,''
    (Cambridge University Press, 1997).
\bibitem{bib:fchc} U. Frisch et al., ``Lattice-Gas Hydrodynamics in Two
    and Three Dimensions,'' {\em Complex Syst.} {\bf 1} (1987) 648.
\bibitem{bib:adler} C. Adler, B.M. Boghosian, E.G. Flekk\o y,
    N. Margolus, D.H. Rothman, ``Simulating Three-Dimensional
    Hydrodynamics on a Cellular-Automata Machine,'' {\em J. Stat. Phys.}
    {\bf 81} (1995) 105-128.
\bibitem{bib:rk} D.H. Rothman and J.M. Keller, ``Immiscible
    Cellular-Automaton Fluids,'' {\em Phys. Rev. Lett.} {\bf 56} (1988)
    889.
\bibitem{bib:chen} H. Chen, S. Chen; G.D. Doolen, Y.C. Lee,
    ``Multithermodynamic Phase Lattice-Gas Automata Incorporating
    Interparticle Potentials,'' {\em Phys. Rev. A} {\bf 40} (1989)
    2850--2853.
\bibitem{bib:az} C. Appert and S. Zaleski, ``Lattice Gas with a
    Liquid-Gas Transition,'' {\em Phys. Rev. Lett.} {\bf 64} (1990) 1.
\bibitem{bib:shanchen} X. Shan, H. Chen, ``Lattice Boltzmann Model for
    Simulating Flows with Multiple Phases and Components,'' {\it
    Phys. Rev. E} {\bf 47} (1993) 1815-1819.
\bibitem{bib:yeomans} M.R. Swift, W.R. Osborn, J.M. Yeomans, ``Lattice
    Boltzmann Study of Hydrodynamic Spinodal Decomposition,'' {\it
    Phys. Rev. Lett.} {\bf 75} (1995) 830-833; M.R. Swift,
    S.E. Orlandini, W.R. Osborn, J.M. Yeomans, ``Lattice Boltzmann
    Simulations of Liquid-Gas and Binary Fluid Systems,'' {\it
    Phys. Rev. E} {\bf 54} (1996) 5041-5052.
\bibitem{bib:mdsq} M. Colvin, A. Ladd, B. Alder, ``Maximally Discretized
    Molecular Dynamics,'' {\it Phys. Rev. Lett.} {\bf 61} (1988) 381.
\bibitem{bib:gunn} J. Gunn, C. McCallum, K. Dawson, ``Dynamical
    Lattice-Model Simulation,'' {\it Phys. Rev. E} {\bf 47} (1993)
    3069-3080.
\bibitem{bib:bce} B.M. Boghosian, P. Coveney, and A. Emerton, ``A
    Lattice-Gas Model of Microemulsions,'' {\em Proc. Roy. Soc. A} {\bf
    452} (1996) 1221.
\bibitem{bib:sep} A.N. Emerton, P.V. Coveney, B.M. Boghosian,
    ``Lattice-Gas Simulations of Domain Growth, Saturation and
    Self-Assembly in Immiscible Fluids and Microemulsions,'' {\em
    Phys. Rev. E} {\bf 55} (1997) 708-720.
\bibitem{bib:ifl} F.W. Starr, S.T. Harrington, B.M. Boghosian,
    H.E. Stanley, ``Interface Roughening in a Hydrodynamic Lattice-Gas
    Model with Surfactant,'' {\em Phys. Rev. Lett.} {\bf 77} (1996)
    3363-3366.
\bibitem{bib:stl} A.N. Emerton, F.W.J. Weig, P.V. Coveney,
    B.M. Boghosian, ``Shear Induced Isotropic-to-Lamellar Transition in
    a Lattice-Gas Automaton Model of Microemulsions,'' {\em J. Phys.:
    Cond. Mat.} {\bf 9} (1997) 8893-8905.
\end{thebibliography}
\end{document}